\documentclass[useAMS,usenatbib]{mn2e}

\usepackage{graphicx}
\usepackage{color}
\usepackage{amssymb}
\usepackage{amsmath}
\usepackage{enumitem}
\usepackage[pdfborder={0 0 0}]{hyperref}

\pdfoutput=1

\title[Starbursts in simulated dwarf galaxies]{Gaseous infall triggering starbursts in simulated dwarf galaxies }
\author[R. Verbeke et al.]
  {R.~Verbeke$^1$\thanks{E-mail:
  {Robbert.Verbeke@UGent.be}.}, S.~De Rijcke$^1$\thanks{E-mail:
  {Sven.DeRijcke@UGent.be}.}, M.~Koleva$^1$, A.~Cloet-Osselaer$^1$, \and B.~Vandenbroucke$^1$,
  J.~Schroyen$^1$\\ $^1$Sterrenkundig
  Observatorium, Ghent University, Krijgslaan 281, S9, 9000 Gent,
  Belgium}
\date{Accepted 2014 May 9, Received 2014 April 29; in original form 2014 April 4}

\pagerange{\pageref{firstpage}--\pageref{lastpage}} \pubyear{2014}

\def\LaTeX{L\kern-.36em\raise.3ex\hbox{a}\kern-.15em
    T\kern-.1667em\lower.7ex\hbox{E}\kern-.125emX}

\newcommand{\unit}[1]{\ensuremath{\, \mathrm{#1}}}
\newcommand{\diff}{\mathrm{d}}

\begin{document}
\label{firstpage}

\maketitle

\begin{abstract}
Using computer simulations, we explored gaseous infall as a possible explanation for the starburst phase in Blue Compact Dwarf galaxies. We simulate a cloud impact by merging a spherical gas cloud into an isolated dwarf galaxy. We investigated which conditions were favourable for triggering a burst and found that the orbit and the mass of the gas cloud play an important role. We discuss the metallicity, the kinematical properties, the internal dynamics and the gas, stellar and dark matter distribution of the simulations during a starburst. We find that these are in good agreement with observations and depending on the set-up (e.g. rotation of the host galaxy, radius of the gas cloud), our bursting galaxies can have qualitatively very different properties. Our simulations offer insight in how starbursts start and evolve. Based on this, we propose what postburst dwarf galaxies will look like.

\end{abstract}


\section{Introduction}

Blue Compact Dwarfs (BCDs) are dwarf galaxies which are currently undergoing an intense period of star formation. The exact classification requirements for a dwarf galaxy to be a BCD differ from author to author. \cite{zwicky64} first coined the term \emph{compact}, in order to classify objects that could only just be distinguished from stars on photographic plates. \cite{thuan81} defined a BCD as a galaxy with low luminosity ($M_B \geq -18$ mag), with small optical sizes ($\leq 1$ kpc), and an optical spectrum that exhibits sharp narrow emission lines superposed on a blue continuum, which are both generally caused by a high star formation rate (SFR). This is the reason BCDs were initially thought to be young objects going through their first episode of star formation.

Deeper observations have shown that most BCDs in fact have a large population of old stars \citep{loose86, kunth88, papaderos96a, cairos01}. \cite*{loose86} proposed a classification scheme based on the shape of the starburst region(s) and of the host galaxy.
\begin{itemize}[leftmargin=0.5cm]
\item i0 BCDs show no evidence of having a host galaxy and are candidates of being true young objects.
\item nE BCDs have a nucleated starburst component, embedded in a regular, elliptical host galaxy.
\item iE BCDs also have an elliptical host galaxy, but the starburst component is irregular.
\item iI BCDs have an irregular starburst embedded in an irregular host galaxy. This class can be further subdivided into
\begin{itemize}[label=$\vartriangleright$,leftmargin=0.5cm]
\item iI,C or cometary BCDs, which have very elongated host galaxies, with the star formation happening near one of the ends.
\item iI,M BCDs, which show clear signs of merging.
\end{itemize}
\end{itemize}

Based on surface brightness profiles, \cite{papaderos96b, salzer99} compared the structural parameters of BCD host galaxies to those of other dwarf galaxies and found that BCDs generally have smaller scale lengths and higher central surface brightnesses, meaning that their host galaxies are in general more compact than non-bursting dwarfs. Deeper observations have shown that the fainter, outer regions of BCDs (between 26 and 28 mag/arcsec$^2$) do agree with dE/dIrr data \citep{micheva13a}, although this might not hold for fainter BCDs \citep{micheva13b}.

BCDs generally have a large central concentration of neutral gas \citep[e.g.][]{taylor94,taylor95,vanzee98, simpson00, zhao13, lelli14}, although a variation in column density between different BCDs has been found \citep{simpson11}. It was found that their gas can have very steep rotation curves, suggesting their dark matter is centrally concentrated as well \citep[e.g.][]{vanzee01, lelli12a, lelli12b} .

The gaseous metallicity of BCDs was found to generally be lower than that of dIs and dEs \citep[e.g.][]{terlevich91,izotov94, izotov99,hunter99}. \cite{zhao13} found that nE and iE BCDs are more metal-rich than iI BCDs, suggesting they might be at different evolutionary stages or have different progenitors.

Using an extended version of the {\sc Gadget-2} N-body/SPH code, we can self-consistently simulate isolated dwarf galaxies \citep{valcke08, schroyen11, cloetosselaer12,schroyen13,cloetosselaer14}. In rotating models, the star formation is continuous while in the non-rotating dwarfs, we can find a SFR that goes up and down with a period of $\sim 10^8-10^9$ years \citep[see also][]{stinson07, revaz09}. However, even in this breathing star formation, we can not identify true bursts.

Several external trigger mechanisms for the starburst in BCDs have been proposed and investigated. Tidal interaction could be the trigger \citep{brinks88, brinks90, campos91, noeske01}, although not every BCD has a massive companion \citep{campos93, telles95}. Mergers between gas-rich dwarf galaxies has also been suggested \citep{ostlin01}, which was confirmed as a plausible mechanism to trigger a starburst in simulated galaxies \citep{bekki08,dimatteo97,cloetosselaer14}. 

As not all BCDs show signs of merging or tidal interaction, several internal mechanisms have been proposed as well, which are not present in our simulation code. \cite{elmegreen12} suggested in-spiraling clumps to drive dark matter, stars and gas from the halo towards the centre, which could feed the starburst. The stellar kinematics of BCDs can be very irregular, with the star forming regions standing out as dynamically distinct entities \citep{koleva14}, suggesting they could be experiencing some form of interaction, even when galaxies seem to be in isolation. External gas infall has been suggested based on their high {\sc HI} mass, disturbed and extended {\sc HI} haloes and their low metallicities \citep{gordon81, lopezsanchez12,sanchezalmeida13,zhao13,sanchezalmeida14}. {\sc HI} clouds with no clear optical counterpart have been found around several BCDs \citep{taylor94, hoffman03, thuan04b, ramya09}, so infall of such gas clouds is not an unlikely event.

In this paper, we shall use computer simulations to investigate whether a low metallicity gas cloud can trigger a starburst event in dwarf galaxies. This gas cloud can represent a truly external object or one that was formed within the galaxies halo.

An overview of the simulations we used is presented in \S 2. \S 3 gives an overview of our results, which are discussed in \S 4. We end with a summary in \S 5.

\section{Simulations}
\label{sec:simulations}

A valuable way to understand the underling physics in galaxies is to run N-body/SPH simulations. When studying dwarf in particular, this is even more so since we can reach very high resolution for these low mass objects. To the freely available N-body/SPH code {\sc Gadget-2} \citep{springel05}, we added star formation, gas cooling and heating, chemical enrichment, and feedback \citep{valcke08, valcke10, schroyen11, cloetosselaer12,derijcke13,schroyen13,cloetosselaer14}. The simulations were then analysed using our own made analysis tool {\sc Hyplot}, which is publicily available\protect\footnotemark[1] and was used to produce all figures in this paper.

\footnotetext[1]{\url{http://sourceforge.net/projects/hyplot/}}

\subsection{Initial conditions for the galaxy}

The basic initial set up for our simulations is a halo of dark matter (DM) particles with gas particles added on top. We start the simulation at redshift $z = 4.3$ and let it evolve for a period of $12.22$ Gyr, until $z = 0$. We employ a flat, $\Lambda$-dominated cosmology with cold dark matter, with $h=0.71, \Omega_\mathrm{M} = 0.2383, \Omega_{\mathrm{DM}} = 0.1967, \Omega_{\mathrm{baryon}}/\Omega_{\mathrm{DM}} = 0.2115$. 
\newline

The dark matter halo is given an NFW density profile \citep*{NFW}

\begin{equation}
\label{nfw}
\rho_\mathrm{DM}(r) = \frac{\rho_{s}}{\frac{r}{r_{\mathrm{DM},s}} \left( 1 + \frac{r}{r_{\mathrm{DM},s}} \right)^2}, \hspace{0.5cm}\mathrm{for}\ r < r_\mathrm{DM,max},
\end{equation}
with $\rho_{s}$, $r_{\mathrm{DM},s}$ and $r_\mathrm{\mathrm{DM},max}$ the characteristic density, the scale length and the cut-off radius respectively. Eq. (\ref{nfw}), which gives a cusped profile, was derived from large-scale cosmological, dark matter only simulations. It was found that through gravitational interaction with the baryons, the dark matter halo  will evolve from having a cusp to a core \citep{read05, cloetosselaer12, teyssier13}. Hence, this density profile is representative only for the start of the simulations.

Following \cite{revaz09, schroyen13}, the gas halo is set up with a pseudo-isothermal density profile

\begin{equation}
\label{pseudo-isothermal}
\rho_g(r) = \frac{\rho_{0}}{1+\frac{r^2}{r^2_{g,s}}}, \hspace{0.5cm}\mathrm{for}\ r < r_\mathrm{g,max},
\end{equation}
with $\rho_{0}$, $r_{g,s}$ and $r_\mathrm{g,max}$ again the central density, the scale length and the cut-off radius respectively. The gas is given an initial temperature of $10^4$ K, which is roughly its virial temperature, and starts with zero metallicity. Both haloes are sampled using a Monte Carlo technique.

The gas halo can be given a rotation, which has a profile given by

\begin{equation}
\label{eq:rot_prof}
v_\mathrm{rot}(r) = \frac{2}{\pi}\mathrm{arctan}\left( \frac{r}{r_\mathrm{rot, s}}\right) v_\mathrm{max},
\end{equation}
with $v_\mathrm{max}$ the rotational velocity at high radii and $r_\mathrm{rot, s}$ the scale length of the rotational profile. For all our simulations, we took $r_\mathrm{rot, s} = 1$ kpc. \cite{schroyen11} compared simulations with rotation to those without and found them to be qualitatively very different. The most important finding was that in dwarf galaxies with significant rotation, the star formation is spread out almost homogeneously over the galactic body, as opposed to being localized to the centre. 
\newline

The bulk of our dwarf galaxy models have a dark matter halo of $2.5 \cdot 10^9 \unit{M}_\odot$ and an intial gas mass of $5.2\cdot 10^8 \unit{M}_\odot$ ("low mass dwarf galaxy models"). This will result in a galaxy with $-14$ mag $\lesssim M_B \lesssim -12$ mag at redshift $z = 0$, where models with higher rotation are more luminous. To see how our results translate to higher luminosities, we also set up galaxies with dark matter mass $10^{10} \unit{M}_\odot$ and initial gas mass $2\cdot 10^{9} M_\odot$ ("high mass dwarf galaxy models"), which will result in galaxies with $M_B \sim -15$ mag. 

We used $2\cdot 10^5$ dark matter and $2\cdot 10^5$ gas particles for the low mass models, resulting in dark matter and gas particles with masses of $\sim 12.5\cdot 10^3 \unit{M}_\odot$ and $\sim 2.6 \cdot 10^3 \unit{M}_\odot$, respectively. 800000 of each particle type were used for the high mass models, giving roughly the same mass resolution.

\begin{table*}
\begin{minipage}{\textwidth}
\begin{center}
\caption{Colums indicate (1) model name, (2) dark matter mass of the host galaxy, (3) gas mass of the host galaxy, (4) the maximal rotational velocity, (5) the mass of the gas cloud, (6) the type of gas cloud (E - extended or C - compact), (7) the radius of the gas cloud, (8) the time the gas cloud is introduced to the system, (9) the initial position of the gas cloud, (10) the initial velocity of the gas cloud, (11) the peak SFR for the merger simulation, (12) the burst factor (the peak in SFR divided by the SFR of the host galaxy in isolation, averaged over 3 Gyr.}
\begin{tabular}{lrrrrrrrrrrr}
\hline

Name		&	$M_\mathrm{host, DM}$		&	$M_\mathrm{host, g}$		&	$v_\mathrm{max}$		&	$M_\mathrm{GC}$	& E/CGC & 	$R_{GC}$		&	$t_\mathrm{init}$	&	($x,y,z$)	&	($v_x,v_y,v_z$)	&	SFR$_\mathrm{peak}$	&	$b$\\
		&	[$10^{10}\unit{M}_\odot$]	&	[$10^{10}\unit{M}_\odot$]	&	[km/s]					&	[$10^{6}\unit{M}_\odot$]	&	&	[kpc]			&	[Gyr]						&	[kpc]		&	[km/s]			&	[$10^{-2}\unit{M}_\odot/$yr]	\\ \hline
B1		&	0.25		&	0.052	&	1.0	&	52	&	E	&	7.5	&	10.33	&	(-1,30,0)	&	(0,-25,0)	&	1.9	&	17.67	\\
B2		&	0.25		&	0.052	&	1.0	&	42	&	C	&	2.0	&	10.33	&	(-1,30,0)	&	(0,-25,0)	&	1.1	&	9.73		\\
B3		&	0.25		&	0.052	&	5.0	&	52	&	E	&	7.5	&	9.39		&	(-2,30,0)	&	(0,-25,0)	&	1.8	&	6.54		\\
B4		&	0.25		&	0.052	&	5.0	&	42	&	C	&	2.0	&	9.54		&	(-2,30,0)	&	(0,-35,0)	&	3.0	&	10.71	\\
B5		&	1		&	0.21		&	2.5	&	52	&	E	&	7.5	&	10.03	&	(-1,30,0)	&	(0,-45,0)	&	3.6	&	3.81		\\
B6		&	1		&	0.21		&	2.5	&	42	&	C	&	3.0	&	10.13	&	(-2,30,0)	&	(0,-20,0)	&	4.9	&	5.16		\\	

\hline
\end{tabular}
\label{tab:mergers}
\end{center}
\end{minipage}

\end{table*}

\subsection{Astrophysical prescriptions}
Starting from these initial conditions, we can integrate the equations of motions, using our modified version of {\sc Gadget-2}.
\subsubsection{Radiative cooling}

\cite{derijcke13} calculated self-consistent metallicity dependent cooling curves ranging from $T=10$ K to $T=10^9$ K. These are dependent on temperature, density, [Fe/H], and [Mg/Fe], which serve as tracers for the chemical enrichment by supernovae (SN) type Ia and type II, respectively. No UV background heating was concluded in the cooling curves we used.

\subsubsection{Star formation}
As the gas cools, it fragments and collapses to dense regions and is allowed to form stars. At each time step, we check the following conditions for each gas particle

\begin{align}
&\vec{\nabla}\cdot \vec{v} < 0, & &\mathrm{(Convergence\ criterium)},  \label{eq:convergence_crit}\\ 
&T < T_{\mathrm{crit}} = 15000 \unit{K}, & &\mathrm{(Temperature\ criterium)}, \label{eq:temp_crit} \\
&\rho_{\mathrm{g}} < \rho_{\mathrm{crit}}, & &\mathrm{(Density\ criterium)}. \label{eq:density_crit}
\end{align}

The convergence criterium (\ref{eq:convergence_crit}) expresses that the gas is collapsing, the temperature criterium (\ref{eq:temp_crit}) expresses that the gas particle can not have a temperature that is too high and the density criterium (\ref{eq:density_crit}) ensures that the gas is dense enough. For $\rho_{\mathrm{crit}}$, the critical density, we use a high density threshold ($\rho_{\mathrm{crit}} = 100\ \mathrm{amu\ cm}^{-3}$) as it has been shown to provide a better prescription for star forming regions \citep{governato10, schroyen13}. The density criterium is the strictest one, as gaseous regions will be unlikely to reach the critical density if they are too warm or not collapsing.

If a gas particle satisfies conditions (\ref{eq:convergence_crit})-(\ref{eq:density_crit}), it is elligble to become a stellar particle, representing a single-age, single-metallicity stellar population. The actual star formation is implemented by a Schmidt law \citep{schmidt59}
\begin{equation}
\frac{\diff\rho_{*}}{\diff t} = -\frac{\diff\rho_{\mathrm{g}}}{\diff t} = c_{*} \frac{\rho_{\mathrm{g}}}{t_{\mathrm{g}}},
\end{equation}
with $\rho_{*}$ and $\rho_\mathrm{g}$ respectively the stellar and gas density, $c_* =0.25$ the dimensionless star formation efficiency and $t_{\mathrm{g}} = 1/\sqrt{4\pi G\rho_\mathrm{g}}$ the dynamical time scale. 

\subsubsection{Feedback and chemical enrichment}
The newly formed stellar particles are modelled with a Salpeter initial-mass function \citep{salpeter55}. Young stellar particles give (thermal) feedback due to supernovae type II and stellar winds, while a delay of 1.5 Gyr is employed for supernovae type Ia feedback to start \citep{yoshii96}. The energy output of both types of supernovae is $10^{51}$ ergs and of stellar winds $10^{50}$ ergs. The actual energy absorbed by the interstellar medium is determined by the feedback effiency $\epsilon_{\mathrm{FB}}$. To produce sensible results, a high feedback efficiency is needed when using a high density threshold and no UV background \citep{cloetosselaer12}. We chose $\epsilon_{\mathrm{FB}} = 0.7$. Along with the thermal energy from feedback, the gas particles are also enriched with metals produced in the stars and in supernovae. 

As our simulations are not able to resolve the hot, low density cavities blown by supernovae, a gas particle that is receiving stellar feedback is not allowed to cool radiatively during that time step.

\subsection{Gas clouds}
\label{sec:gasclouds}

To simulate a merger with a gas cloud, we first simulate a dwarf
galaxy in isolation. Then, we introduce a gas cloud to the system at a
certain time, a certain distance and with a certain velocity. We let
it evolve until $z = 0$, allowing us to compare with observed BCDs at
low redshift. While metallicity data exists for Milky Way HVCs,
  it is, at least to our knowledge, non-existent for gas clouds
  surrounding BCDs. Since BCDs are found predominantly in low-density
  environments such as the field, one would not expect gas clouds to
  have been significantly polluted. Therefore, the gas in the
  infalling gas cloud is initially given zero metallicity.

\subsubsection{Mass}

{\sc HI} clouds with masses up to $\sim 10^7 \unit{M}_\odot$ were
found around BCDs \citep{hoffman03, thuan04b}. For more estimates for
the masses of infalling gas clouds, we could take a look at the masses
of the high velocity clouds (HVCs) around the Milky Way and other
giant spirals as more data is available for these
\citep[e.g.][]{oosterloo07,lehner09,putman09}. The HI masses of these
HVCs are estimated to range between $10^4$ M$_\odot$ and $10^7$
M$_\odot$, but with great uncertainties on these masses because of the
uncertain distances \citep{vanwoerden04}. A HVC with HI mass around
$10^8$ M$_\odot$ has been found around M101, which most likely has a
galactic origin \citep{vanderhulst88}, so unlikely to be found around
BCDs.

\cite{keres2009} performed computer simulations of the formation of
these HVCs around a Milky Way sized halo. These produced gas clouds
with total gas masses (both neutral and ionized) ranging from $10^6$
M$_\odot$ to several $10^7$ M$_\odot$, with the lower limit being an
artefact of the low resolution of the simulation. The clouds in this
simulation located at high distances from the galactic centre have a
trend to also be the more massive ones.

Therefore, although a mass of $\sim 10^7$ M$_\odot$ is,
  admittedly, at the high end of the gas cloud mass distribution \citep{pisano07}, such
  massive clouds do exist. We expect that gas clouds with higher
masses will have a bigger influence, so based on these observations
and simulations, we will only discuss gas clouds with $M \approx 10^7
\unit{M}_\odot$.

The density of the gas cloud is chosen to have a $1/r$-profile.

\subsubsection{Radius}

We investigated several radii for the gas cloud and separate the discussion qualitatively in compact gas clouds (CGCs) and extended gas clouds (EGCs). Following the Lema\^itre-Tolman-Bondimodel for a spherical overdensity, we use for EGCs, 

\begin{equation}
\label{eq:R555}
r_{\mathrm{max}} = \frac{0.09617}{5.55^{1/3}} \left[ \frac{M_\mathrm{tot}}{h^2(\Omega_m(1+z)^3+1-\Omega_m)}\right]^{1/3} \mathrm{kpc},
\end{equation} 

with $M_\mathrm{tot}$ the total mass of the gas cloud in M$_\odot$, $h = 0.71$, $z = 4.3$, and $\Omega_m = 0.2383$. E.g. for a gas cloud with mass $5.2\cdot 10^7$ M$_\odot$, Eq. (\ref{eq:R555}) gives us $r_{\mathrm{max}} \approx 7.5$ kpc.

The radius of a CGC we chose ad hoc, but significantly smaller than that obtained from Eq. (\ref{eq:R555}). On the other hand, we also take it large enough to prevent that the density is too high and the gas cloud starts forming stars on its own. For CGCs we typically chose a radius of $2-3$ kpc.

\subsubsection{Temperature}

As an extra condition, we want the gas cloud to remain stable over
several Gyr, so that it does not start collapsing or expanding
significantly. To do so, we set the temperature to be around the
virial temperature
\begin{equation}
\label{eq:vir_temp}
T_{vir} \simeq 6.8 \cdot 10^{-5} \mu  \frac{M_\mathrm{tot}}{r_\mathrm{max}} \frac{\unit{kpc}}{\unit{M}_\odot} \unit{K},
\end{equation}
with $\mu$ the mean molecular weight of the gas (for our purposes $\mu
\approx 1.2$). We evolved isolated gas clouds during one Hubble
  time and found that the density profile within the inner $\sim
  5$~kpc hardly changes. Only the tenuous outer layers expand because
  the gas clouds are set up with a vacuum boundary. For the actual
  simulations, we first let each gas cloud relax for 1 Gyr before
  putting them together with a dwarf galaxy in a merger simulation.

\subsubsection{Velocity}

The initial velocity given to the gas cloud was inspired by the escape velocity of the target galaxy, so the gas cloud seemingly falls in with an initial velocity

\begin{equation}
\label{eq:v_esc}
v_\mathrm{esc} (R) = \sqrt{\frac{2GM}{R}}.
\end{equation}
We put the centre of the gas cloud at $R= 30$ kpc, far enough from the galaxy. Looking into our dwarf galaxy simulations for the total mass enclosed with a radius of 30 kpc, we find $M_{<30\mathrm{kpc}} \approx 2.5\cdot 10^9 \unit{M}_\odot$ for the low mass models and $M_{<30\mathrm{kpc}} \approx 7\cdot  10^9 \unit{M}_\odot$ for the high mass models, which gives us $v_\mathrm{esc} \approx 25$ km/s and $v_\mathrm{esc} \approx 45$ km/s, respectively.

\subsection{The models}
\label{sec:models}
Almost 150 simulations were run, all with variations in their set-up. The parameters that yield the biggest qualitative difference in the produced burst are the rotation rate of the host galaxy, the mass of the host and the radius of the gas cloud falling in.
In this paper, we will explicitly discuss 6 different simulations, in which these effects are very clear. 
\begin{itemize}[leftmargin=0.5cm]
\item B1 is a low-mass, slowly rotating host galaxy onto which an EGC falls in.
\item B2 has the same host as B1, but a CGC is falling in.
\item B3 is a low-mass host with fast rotation and an EGC falling in.
\item B4 has the same host galaxy as B3, but with a CGC falling in.
\item B5 is a high mass host and an EGC falling in.
\item B6 has a high mass host galaxy, but with a CGC falling in.
\end{itemize}
Tabel \ref{tab:mergers} gives a more detailed overview of the properties of the different models.

Other variations were explored, like the initial velocity of the gas cloud or the trajectory along which it falls in. These are not discussed explicitly using different models, but their results will be discussed where relevant in \S \ref{sec:results}.

\section{Results}
\label{sec:results}

\subsection{Statistics}
\label{sec:statistics}
Before looking at the models presented in \S \ref{sec:models} and Table \ref{tab:mergers}, we will discuss some general statistics of our simulations. For this, we define the burstfactor $b$ as the peak SFR during the burst divided by the SFR of the isolated host averaged over the last 3 Gyr:

\begin{equation}
\label{eq:burstfactor}
b = \mathrm{SFR}_\mathrm{peak}/\ \overline{\mathrm{SFR}}_\mathrm{host}
\end{equation}

We will identify a burst if $b > 5$ and a strong burst if $b > 10$.

In total, 143 simulations were run. Out of these, 73 ($\sim 50 \%$) had a burst, of which 9 were strong bursts. However, we can divide our simulations in several groups depending on the set-up, to distinguish several determining effects. 

17 of the total simulation sample used a gas cloud with $M < 10^7 \unit{M}_\odot$, of which none produced a burst.

In 8 of the remaining simulations, the gas cloud fell in along a prograde orbit, while in the others a retrograde orbit or a face-on collision was used. In only 1 of the former, a burst could be found. After close inspection of this simulation, we found that the reason for the burst was that the gas cloud already had a high density on its own, out of which a large amount of stars could be formed when the gas cloud was moving through a deep gravitational potential and a high density region.

Of the remaining 118 simulations, 22 were high mass models and 96 were low mass models. 5 ($\sim 20 \%$) of the high mass models had a burst (of which none were strong bursts), while 67 ($\sim 70 \%$) of the low mass models had a burst (out of which 8 were strong bursts).

We do not wish to over-interpret these statistics, as they are not independent (e.g. several share the same host galaxy), but we can draw some general conclusions. 

\begin{itemize}[leftmargin=0.5cm]
\item We need a gas cloud with $M \gtrsim 10^7 \unit{M}_\odot$ to trigger a burst in our models.
\item Gas clouds falling in on a prograde orbit are unlikely to trigger a burst.
\item For a gas cloud with the same mass, higher mass galaxies will be less likely to undergo a burst.
\end{itemize}

\subsection{Gaseous infall and starburst modes}

\begin{figure}
\includegraphics[width=0.47\textwidth]{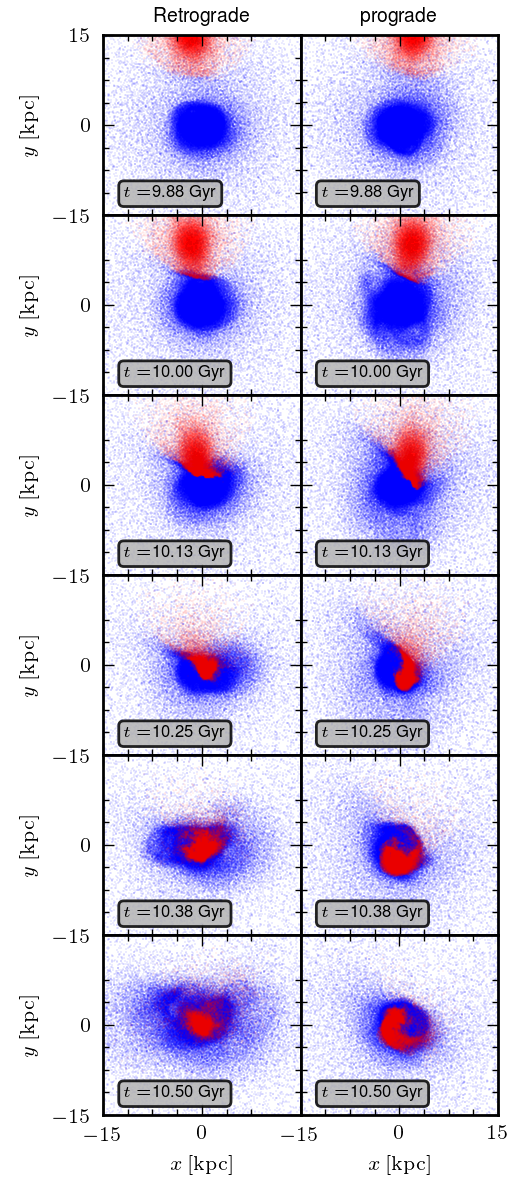}
\caption{{Gaseous infall on a dwarf galaxy. Black points (red in the online colour version) show the gas cloud falling in, gray points (blue in the online colour version) represent the gas of the dwarf galaxy. The left panels show the set-up of B3: a gas cloud falling in on a dwarf galaxy along a retrograde trajectory. The right panel shows the same set-up, but with the initial position of the gas cloud at $(2, 30, 0)$, resulting in a prograde trajectory.}
}\label{fig:DG_GC_merger}
\end{figure}

\begin{figure}
\includegraphics[width=0.47\textwidth]{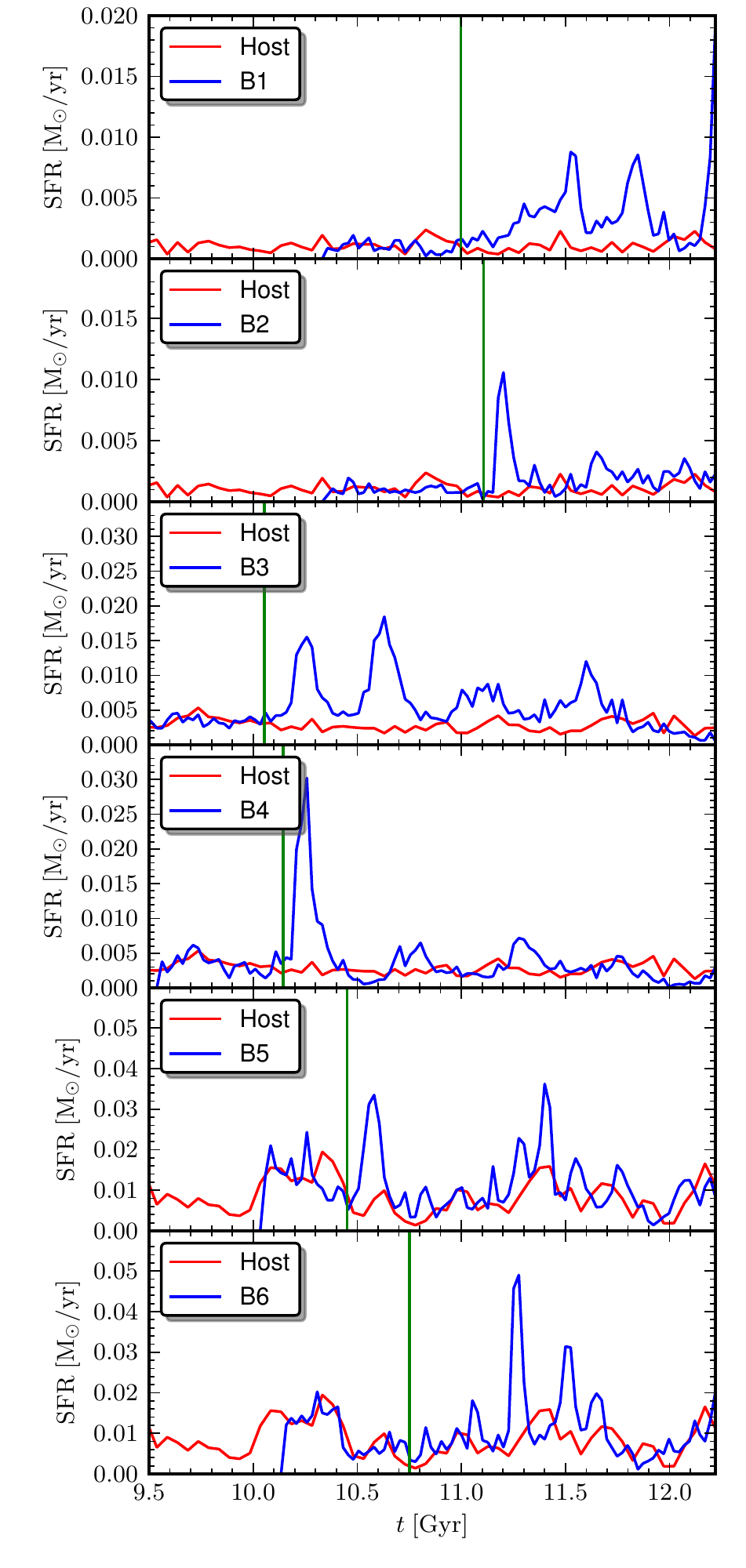}
\caption{Star formation rate of bursting galaxies (black, blue in the online colour version), compared to that of their host galaxy in isolation (gray, red in the online colour version). The models are the ones from Table \ref{tab:mergers}. Vertical lines roughly indicate when the gas cloud falls in on the dwarf galaxy.}
\label{fig:SFR}
\end{figure}

Gaseous infall can trigger enhanced star formation in a galaxy in several ways. This can be seen from Figure \ref{fig:SFR}, where we show the star formation rate in function of time of our six models (black, blue in the online colour version), compared to that of their host galaxies in isolation (gray, red in the online colour version). The vertical lines roughly show the moments the gas cloud enters the system.

When a very extended gas cloud is falling in on a slowly or non-rotating host galaxy, it can compress the gas of the host out of which stars will be born (B5). However, this will not necessarily happen and we might not find a starburst initially. However, all the gas of the host galaxy will be affected, making it more turbulent and more likely to collapse, resulting in a strong burst at later times (B1).

In the case of a rotating galaxy (B3), the story is mainly the same, except that the direction the gas cloud is falling in from is important as it was concluded in \S \ref{sec:statistics}. In a
retrograde merger, the two gas reservoirs collide and lots of gas is driven towards the galaxy center. This is clearly in the left column of Figure \ref{fig:DG_GC_merger}. In a prograde merger, the velocities of the two gas reservoirs closely match and the infalling gas cloud swirls around the
galaxy center. There is no significant angular momentum exchange between both gas reservoirs, no gas is driven towards the galaxy center, and no starburst is ignited. Movies of this can be found online\protect\footnotemark[2].

\footnotetext[2]{Youtube channel of our department: \url{https://www.youtube.com/user/AstroUGent}. A playlist with movies concerning this paper: \url{http://www.youtube.com/playlist?list=PL-DZsb1G8F\_kNcMwgkFNhQpHXRHbOnCH5}}

For the high-mass models, these results remain valid, but since we use a gas cloud of the same mass as in the low mass models, the gas of the galaxy will be less disturbed in this case.

\begin{figure}
\includegraphics[width=0.47\textwidth]{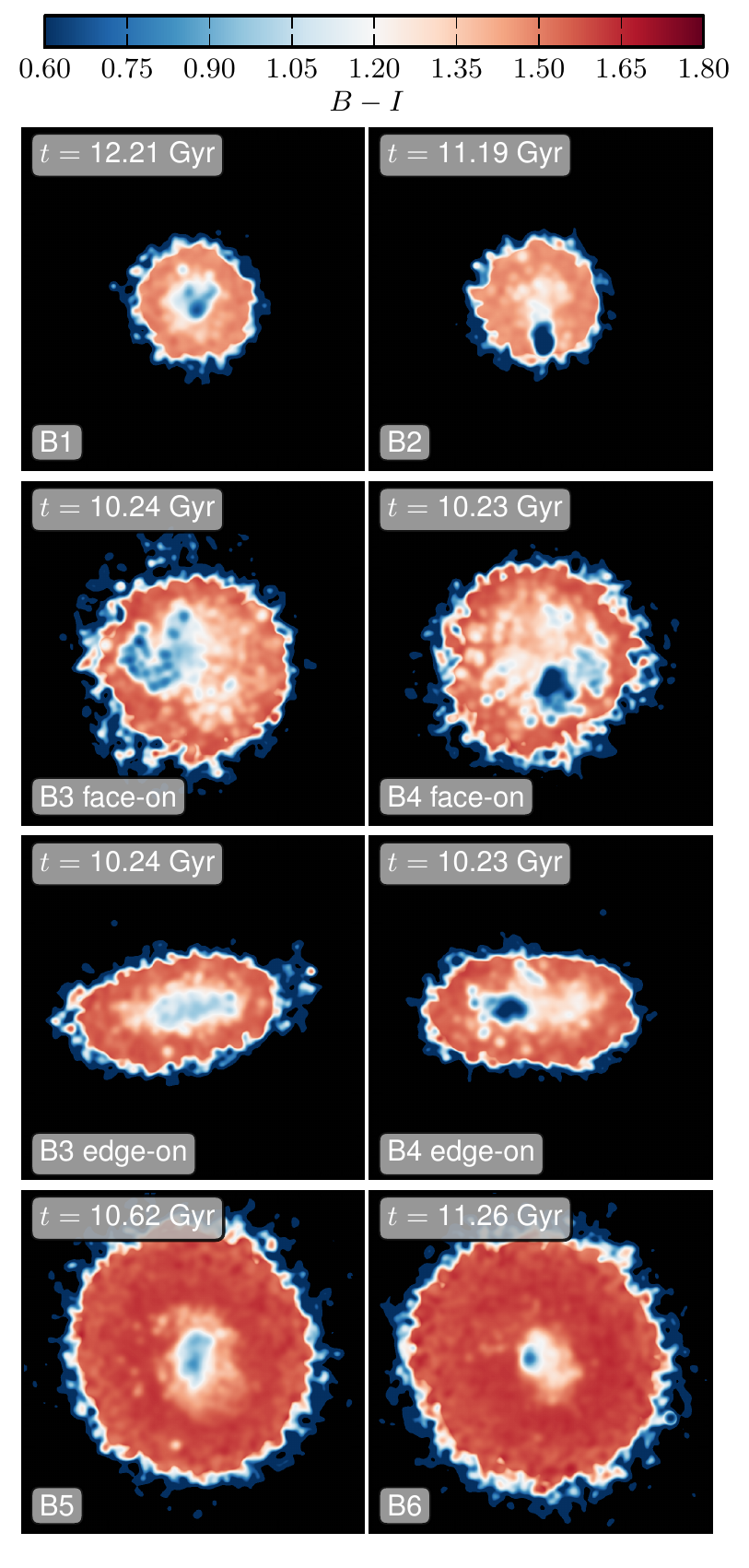}
\caption{$B-I$ colour maps of our models during their bursts. All figures show 10 kpc in both directions.}
\label{fig:colorMaps}
\end{figure}

On the other hand, when we use a compact gas cloud, the gas cloud will only have an effect very locally. When stars are being formed as a direct consequence of the gas cloud impact, they will be formed along the path of the gas cloud. This is the case in B2 and B4. In the absence of rotation, the host galaxy will not be able to efficiently stop the gas cloud, so star forming region will have an elongated shape. Rotating galaxies will be able to stop it more efficiently, resulting in a more centralized star forming region.

In B6, the CGC did not immediately trigger an increased star formation rate. Rather, most of the gas flew through the galaxy, since the gas cloud has been significantly accelerated by the high mass dwarf galaxy. Shortly after however, the gas falls back, triggering the burst. The starburst is more centralized compared to B2 and B4 since the gas falling back in now has low densities. Thus the stars will be formed out of the old gas of the galaxy, much like the simulations with EGCs.

Figure \ref{fig:colorMaps} shows $B-I$ colour images of our models, illustrating these effects. These images were obtained by setting up Cartesian luminosity grids in both bands, to which we applied a Gaussian filter (with $\sigma = 100$ pc) to produce smoother images. The surface brightness of grid cells fainter than 29 mag/arcsec$^2$ were put to this magnitude. The blue edges are an artefact of this truncation, since the B-band will be rounded off more quickly than the I-band, so $B-I$ is underestimated.

\subsection{Duration and periodicity of the starburst}

After the initial burst, feedback in the form of SNII and stellar winds will efficiently remove the central gas reservoir and shut down the burst. Our bursts have lifetimes of a couple of 100 Myr, with a trend of stronger bursts having shorter lifetimes since the feedback will be more intense and remove the gas more efficiently. Indeed, in simulation B1 (top panel of Figure \ref{fig:SFR}) we can identify a period of high star formation lasting $\sim 1$ Gyr, shortly followed by a strong but short burst.

After the burst, we have little gas in the centre of the galaxy, making it easy for the gas to fall back in and produce secondary bursts. We find these subsequent bursts in simulations B1, B3, B5 and B6, as can be seen in Figure \ref{fig:SFR}.

\subsection{Strength of the burst}

The mass of the gas cloud plays an important role in the strength of the burst. The more massive it is, the stronger the burst it can trigger. A gas cloud with $M = 5\cdot 10^7 \unit{M}_\odot$ can temporarily increase the SFR with a factor 20 in the low mass dwarf galaxy models. 

For gas clouds of the same mass, the increase in SFR will be less significant in more massive dwarfs. These dwarf galaxies are already forming stars at higher rates because of the higher gas content and deeper gravitational potential, making it harder to get a relative increase in SFR similar to that in the less massive dwarf galaxy models. The more massive dwarfs will still have an increased SFR during the burst.

The peak in SFR of each model and the burstfactor $b$, which is the relative increase compared to the average SFR of the host over the last 3 Gyr, are given in column 11 and 12 in Table \ref{tab:mergers}.

Secondary bursts can be stronger than the initial burst (e.g. B1) but generally, we can expect the strength of each subsequent burst to be lower than the previous one.

\subsection{Metallicity}
\label{sec:metal}

\begin{figure}
\includegraphics[width=0.47\textwidth]{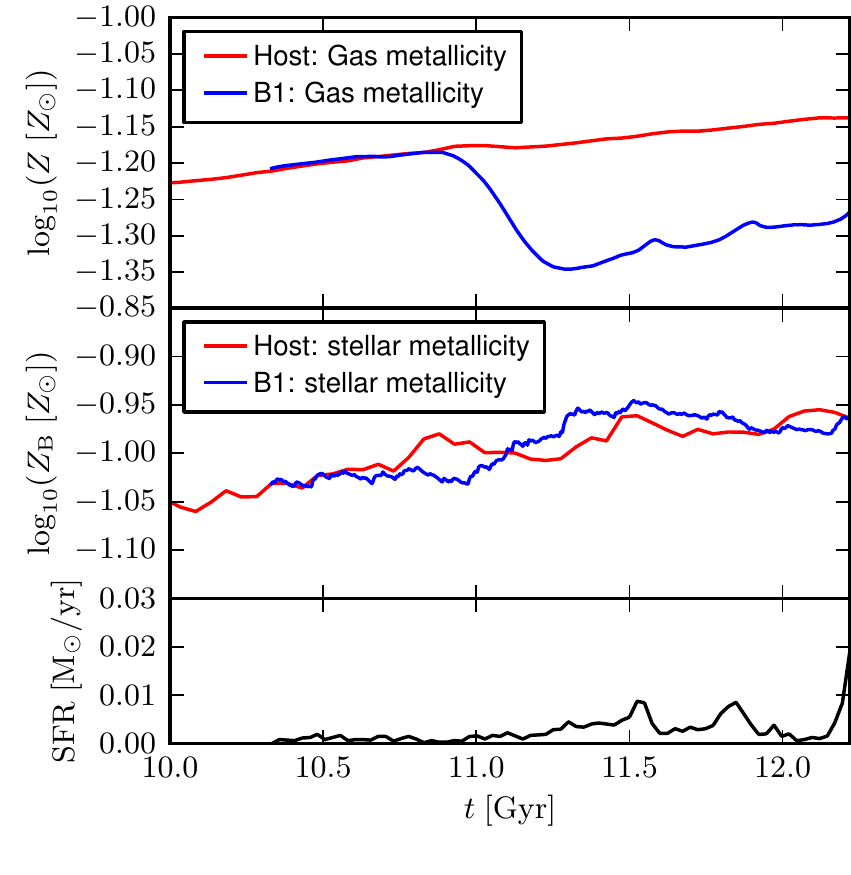}
\caption{Evolution of the metallicity of the gas within 5 kpc (top panel), the B-band luminosity weighted metallicity of the stars (middle panel) and the SFR (bottom panel) of simulation B1. Gray (red in the online colour version) shows the model in isolation, black (blue in the online colour version) shows what happens when a gas cloud falls in.}
\label{fig:metal_evo_B1}
\end{figure}

\begin{figure}
\includegraphics[width=0.47\textwidth]{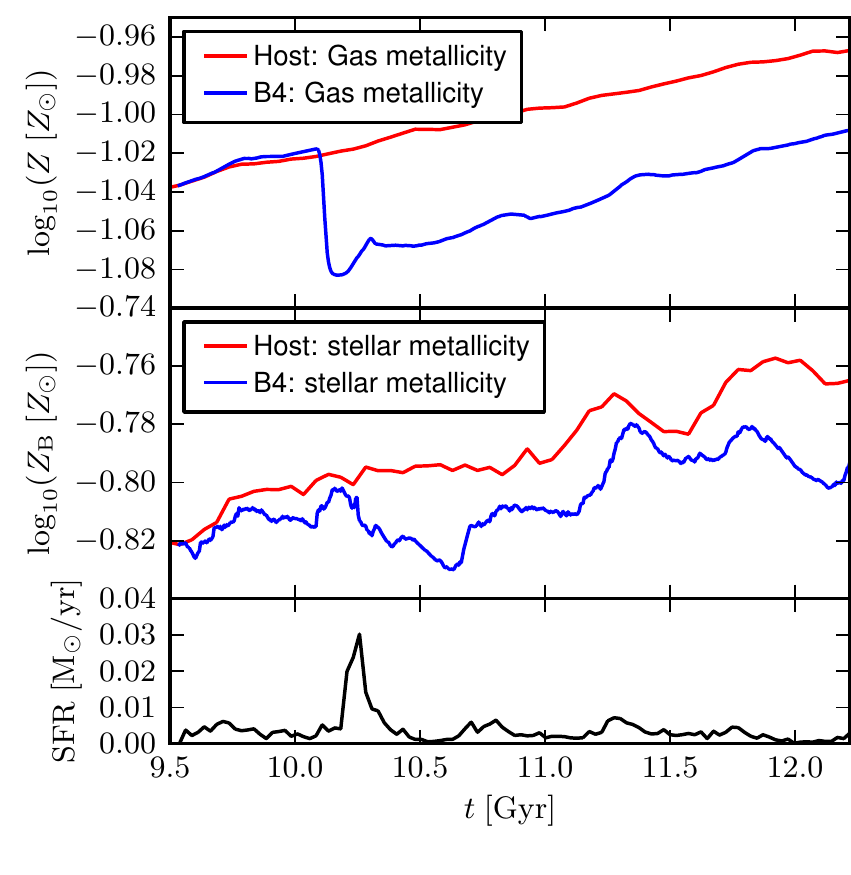}
\caption{Same as Figure \ref{fig:metal_evo_B1} but now for simulation B4.}
\label{fig:metal_evo_B4}
\end{figure}

It is to be expected that the metallicity of the dwarf galaxy will drop when we let a metal-poor gas cloud merge with it. In all our models, we found that the metallicity of the gas within 5 kpc of the dwarf galaxy centre indeed drops significantly when the gas cloud enters the system. We show the evolution of the gaseous metallicity in the top panels of Figure \ref{fig:metal_evo_B1} and Figure \ref{fig:metal_evo_B4} for B1 and B4, respectively.

However, looking at the B-band luminosity weighted metallicity of the stars (middle panel of Figure \ref{fig:metal_evo_B1}), we see that the metallicity can increase significantly following the merger. In fact, the only ones of the models discussed here where the metallicity drops are model B2 and B4, as can be seen in Figure \ref{fig:metal_evo_B4} for the latter. 

In the other models, the gas falling in will not reach high enough densities to form stars at initial impact so if there is an immediate starburst (e.g. B5), the stars will be formed out of the old, metal enriched gas. In the case of bursts arising from the disturbed gas (e.g. B1), the gas will be well mixed and stars will be formed out of metal enriched gas as well. In B2 and B4, on the contrary, a significant part of the starburst is consuming the new, metal-poor gas.

\subsection{Colour}
\label{sec:color}

\begin{figure}

\includegraphics[width=0.47\textwidth]{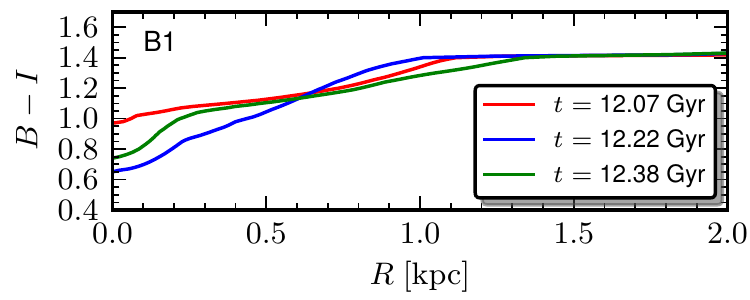}
\includegraphics[width=0.47\textwidth]{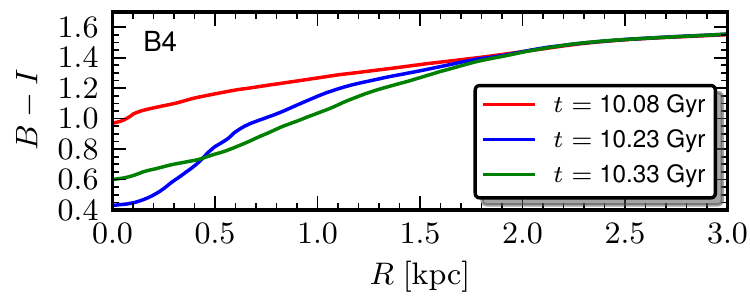}
\caption{$B-I$ colour profile of model B1 (top) and B4 (bottom) before (light gray, red in the online colour version), during (black, blue in the online colour version) and after (gray, green in the online colour version) their respective bursts.}
\label{fig:colorProfile}
\end{figure}

Given the presence of irregular star formation knots in our galaxies during the starburst, we obtain colour profiles using a technique similar to \emph{isophotal integration} for obtaining surface brightness profiles \citep{papaderos96a, micheva13a,micheva13b}. We start from the Cartesian grids used to produce the $B-I$ colourmaps in Figure \ref{fig:colorMaps}. For each colour value, we determine the number of grid cells with colours bluer than this value and add their areas together. These areas are then converted to the equivalent radius of a circle with the same area. This method gives the radius in function of colour and it will produce a monotonically rising profile. The employed technique comes with the caveat that at some point, grid cells containing no stellar particles start influencing the profile. We cut off our profile before this happens. Also, the obtained radius can not be interpreted as a physical distance from the centre of the galaxy, unless in regular circular or elliptical systems. 

The result for B1 and B4 can be seen in Figure \ref{fig:colorProfile}, in which the colour profiles before (light gray, red in the online colour version), during (black, blue in the online colour version) and after (gray, green in the online colour version) their respective bursts are shown. We note that B1 has a strong burst right at the end of the simulation (top panel of Figure \ref{fig:SFR}). To investigate what happens after the burst, we continued this simulation until $t=12.7$ Gyr. For this simulation, we see a clear drop in colour for $R \lesssim 1$ kpc during the burst, with colour difference of over 0.3 mag in the centre, compared to before the burst. After the burst, the galaxy has become redder in the centre than during the burst, but is still bluer than before the burst. For $R \gtrsim 0.5$ kpc, the post-burst galaxy is bluer than during the burst. This can be due to recently formed stars moving to higher radii or stars being formed by the intense feedback following the central starburst. Since the stars are being formed from the metal-enriched gas of the dwarf galaxy, the bluer colours can be solely attributed by the age of the stellar population. For B4, the trend is the same, but we find much bluer colours: $B-I$ drops by $\sim 0.6$ mag in the centre, compared to before the burst. The reason is that now the stars are being formed from the metal-poor gas of the gas cloud.

\subsection{Density profiles}
\label{sec:density_profiles}

\begin{figure}
\includegraphics[width=0.47\textwidth]{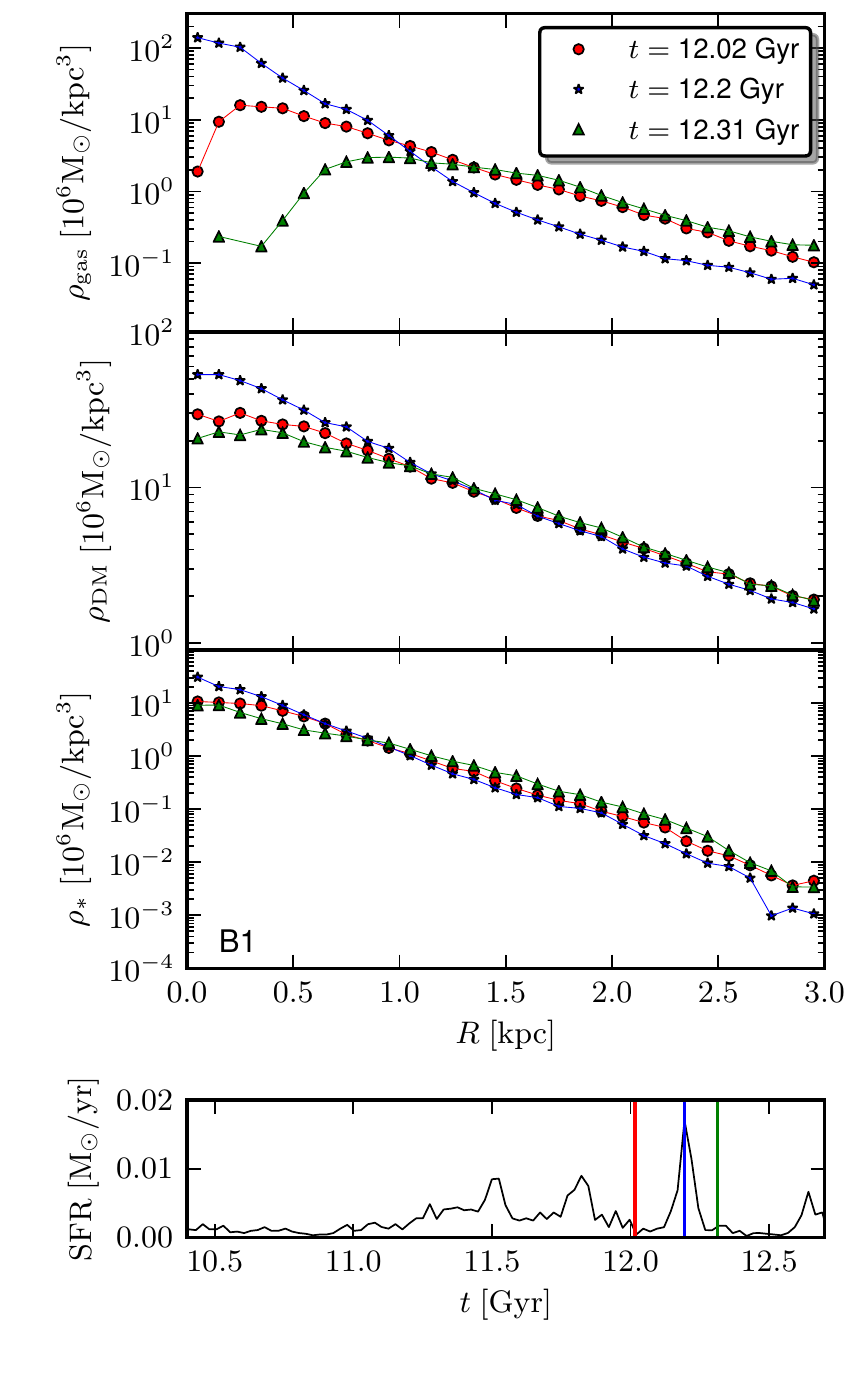}
\caption{Gas, dark matter and stellar density profile of B1 before (light gray circles, red in the online colour version), during (black stars, blue in the online colour version) and after the starburst event (gray triangles, green in the online colour version).}
\label{fig:dens_all_2221}
\end{figure}

Figure \ref{fig:dens_all_2221} shows the density profile of the gas, dark matter and stars, at times before (light gray circles, red in the online colour version), during (black stars, blue in the online colour version) and after the burst (gray triangles, green in the online colour version). We see that for all the components, the central density rises significantly during the burst compared to before, while the density at higher radii drops. After the burst, the central density drops, even below the density before the burst, and gets higher in the outer regions. This effect is strongest for the gas.

No matter how the starburst is triggered exactly, the reason we have a burst is because the gas reaches a high density in a certain area. So the higher central gaseous densities should not come as a surprise. However, this large central gaseous mass causes the rest of the galaxy to contract. Hence we get a more compact galaxy, with a higher concentration of gas, dark matter and stars in the centre (also because of the high star formation) and a lower concentration in the outskirts of the galaxy. This redistribution of matter can in other words be attributed to the deepening of the gravitational potential. This process occurs in all models, albeit less significant in B2 and B4, where the impact of the CGC is a more local effect. 

Shortly after the burst, the gas will receive a lot of feedback from the young stellar particles and will be blown to larger radii. The gravitational potential becomes shallower and the dark matter and stars also move outwards, resulting in a more diffuse galaxy. This process is present in all models and should be only dependent on the strength of the burst, as this determines the amount of feedback and thus the efficiency of the removal of the gas.

Since the expansion due to feedback will be faster than the previous collapse to the centre, the former will have a larger effect. So we are possibly left with a postburst galaxy that is more diffuse than preburst.

\begin{figure}
\includegraphics[width=0.47\textwidth]{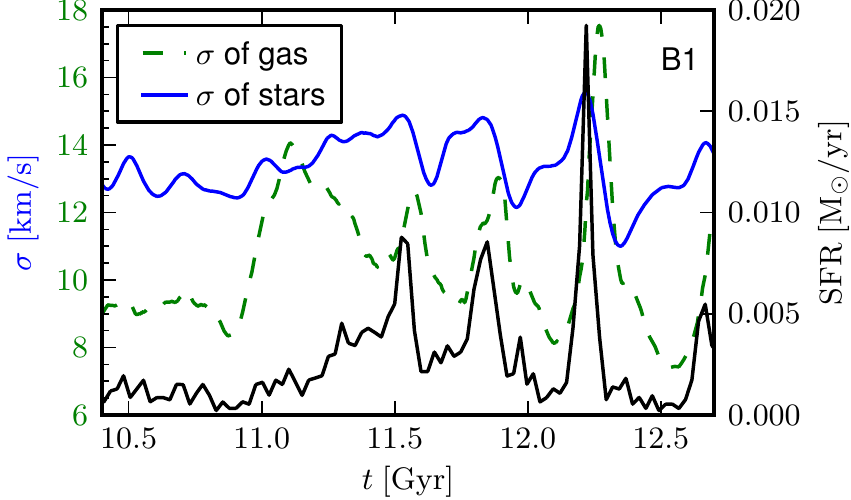}
\caption{Velocity dispersion of the gas (gray dashed line, green in the online colour version) and of the stars (light gray solid line, blue in the online colour version) compared to the SFR (black solid line) for simulation B1.}
\label{fig:velDisp_2221}
\end{figure}

\subsection{Velocity dispersion}
\label{sec:vel_disp}
We have a look at the velocity dispersion of both the gas (within 5 kpc) and the stars, defined as
\begin{equation}
\label{eq:velDisp}
\sigma = \sqrt{\frac{\sigma_x^2+\sigma_y^2+\sigma_z^2}{3}}.
\end{equation}

In Figure \ref{fig:velDisp_2221}, we show the evolution of $\sigma$ of the gas and of the stars of B1, along with the SFR. When the gas enters the system (at $\sim$ 11 Gyr), the velocity dispersion of the gas increases significantely. After it drops back, it follows the general trend of the SFR with a slight delay. The velocity dispersion of the stars increases and decreases together with the SFR. This trend is present in all our models, albeit less significant in the more massive ones B5 and B6.

Before the gas can form stars, it has to cool down and collapse, thus lowering its velocity dispersion. As new stars are being formed, gas with low $\sigma$ is removed and, more importantly, the surrounding gas gets heated and it expands, increasing the gaseous velocity dispersion. This explains why the velocity dispersion of the gas follows the star formation rate with a slight delay.

On the other hand, as the gas collapses and produces a burst, the stellar body also becomes more compact and the velocity dispersion of the stars increases during a starburst. 

Shortly after the burst, the stellar body is more diffuse and the velocity dispersion drops. This explains the concurrence of the stellar velocity dispersion curve and the SFR in Figure \ref{fig:velDisp_2221}. The same is true for the dark matter. As an extra effect, we have a lot of young stars with low $\sigma$, further decreasing the velocity dispersion of the stars.

\subsection{Isophotal maps}
\label{sec:isophotes}

\begin{figure}
\includegraphics[width=0.47\textwidth]{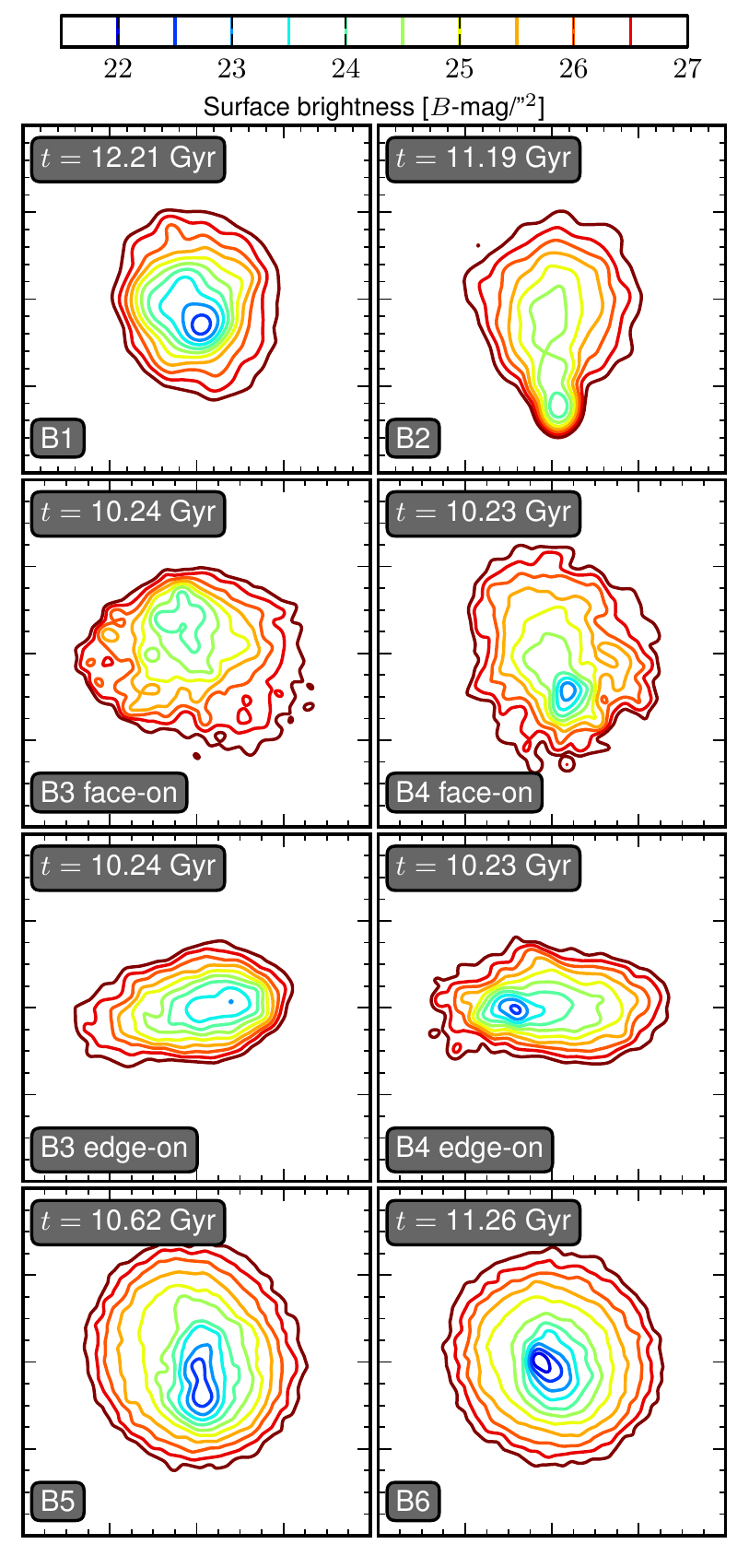}
\caption{Isophotal maps in the B-band of our models during their burst. Sides for B1 and B2 are 4 kpc on both sides, for B3 and B4 6 kpc and for B5 and B6 both sides are 7 kpc. For B3 and B4, the models with a strongly rotating host galaxy, both the edge-on as the face-on maps are shown. Isophotes go from $\mu =$ 27 mag/arcsec$^2$ (most outer and faintest, reddest in the online colour version) up to $\mu =$ 21.5 mag/arcsec$^2$ (most inner and darkest, bluest in the online colour version), in increments of 0.5 mag/arcsec$^2$}
\label{fig:isophotes}
\end{figure}

Figure \ref{fig:isophotes} shows the isophotal maps of our models during their strongest peak in star formation. To obtain these, we produced a cartesian grid (with cells of width 70 pc), calculated the surface brightness in each cell, applied a Gaussian filter ($\sigma = 100$ pc) to smooth the data and plotted the contours. Isophotes go from $\mu =$ 27 mag/arcsec$^2$ (most outer and faintest, reddest in the online colour version) up to $\mu =$ 21.5 mag/arcsec$^2$ (most inner and darkest, bluest in the online colour version), in increments of 0.5 mag/arcsec$^2$

Model B1 has a regular, slightly off-centre star forming region and the outer isophotes are nicely elliptical. Model B2 has a very off-centre star forming region, giving the entire galaxy a very elongated shape. 

For B3 and B4, the models with a host with strong rotation, we show both a face-on and edge-on view. Face-on, both have very irregular outer isophotes, which can be explained by the rotation \citep{schroyen11}. The star forming region in both models is very off-centre. In B3, it is very irregular, while the star forming region in B4 is more regular. However, due to it being off-centre, it also gives the galaxy an elongated shape.

Edge-on, the outer isophotes are flattened and more regular. In B3, the star forming region is off-centre and slightly irregular, while in B4, it is very centralized.

Both the high mass models B5 and B6 show a regular, slightly off-centre star forming region embedded in an elliptical host galaxy.

\section{Discussion}

\begin{figure}
\includegraphics[width=0.47\textwidth]{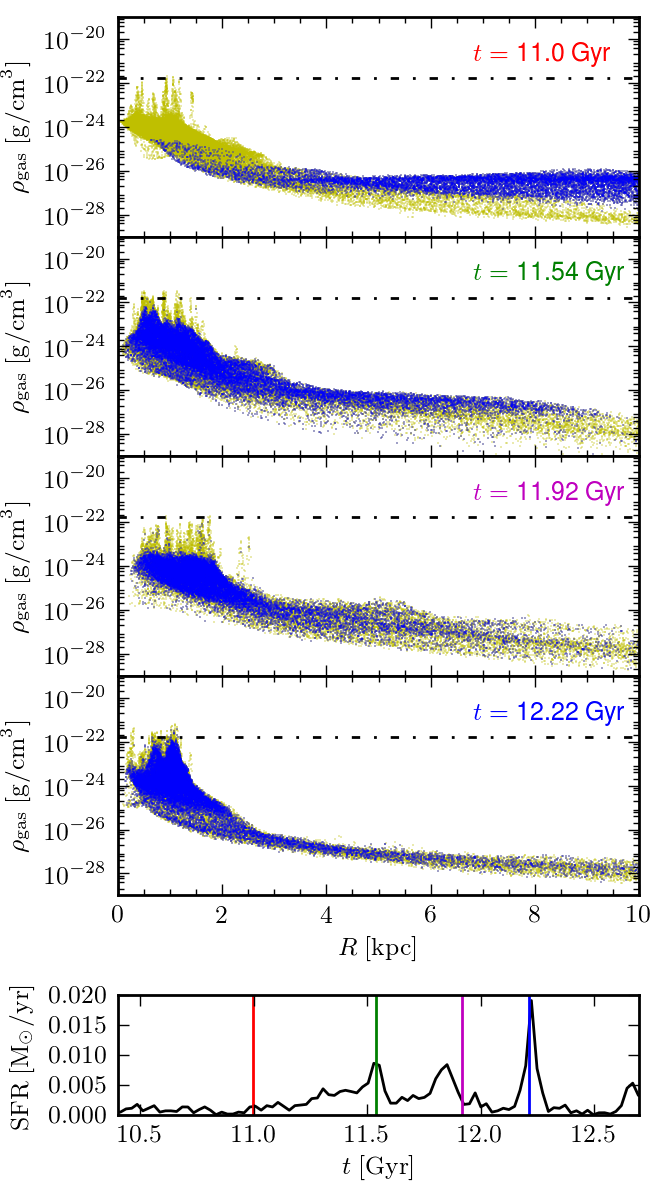}
\caption{Scatter plot of the density and radius of all gas particles of simulation B1. Black points (yellow in the online colour version) represent the old gas of the dwarf galaxy, gray points (blue in the online colour version) show the infalling gas cloud. The dash-dot line shows the density threshold for star formation.\label{fig:densityRadiusGas}}
\end{figure}

\subsection{The triggering mechanism}

Having established that an infalling gas cloud can trigger a
significant increase in the star formation of a dwarf galaxy, we can
wonder what exactly is the reason of this increase. In Figure
\ref{fig:densityRadiusGas}, we show the radius and density of all gas
particles in simulation B1 at different times. The gas already present
in the dwarf galaxy is shown by black dots (yellow in the online colour version), while the gray dots (blue in the online colour version)
represent the newly added gas of the gas cloud. The first panel
($t=11$ Gyr) shows the extended gas cloud falling in. In the second
panel ($t=11.54$ Gyr), the influx of gas has led to the formation of
many star-forming clouds and consequently to an increased
SFR. Remarkably, virtually no ``new'' gas particles are mixed into
these high-density, star-forming regions and the stars are being
formed out of the ``old'' gas of the galaxy. This causes the stellar
metallicity to further increase (\S \ref{sec:metal}). Clearly, this
first sharp increase of the SFR is the direct result of the gas cloud
compressing the old gas of the galaxy. This initial increase in SFR
when the gas cloud enters the system, can be seen in all our models,
except in B6. On the other hand, when the gas cloud is initially dense
enough, stars can form out of the new gas, resulting in a more
metal-poor starburst. This occurs in simulations B2 and B4.

Moreover, we can conclude from our simulations that gas cloud mergers
also have a longer lasting impact on the SFR. The increased turbulence
of the interstellar gas following a merger and starburst lasts for
several Gyrs, as can be seen in the velocity disperion of the gas (\S
\ref{sec:vel_disp}). During this time, the gas appears to be much more
prone to gravitational collapse and therefore to subsequent starburst
events. In the third panel ($t=11.92$ Gyr) of Figure
\ref{fig:densityRadiusGas}, the feedback following the previous star
formation has blown out the gas from the central region, lowering the
SFR. Around $R\sim 5$ kpc, for instance, one can see this gas being blown
out. In the fourth panel ($t = 12.22$ Gyr), the gas has fallen back in
and a starburst ensues, over a gigayear after the merger. Now, the gas
is well mixed and the stars are being formed out of both the old and
the newly added gas.

\subsection{Surface brightness profile}

\begin{figure}
\includegraphics[width=0.47\textwidth]{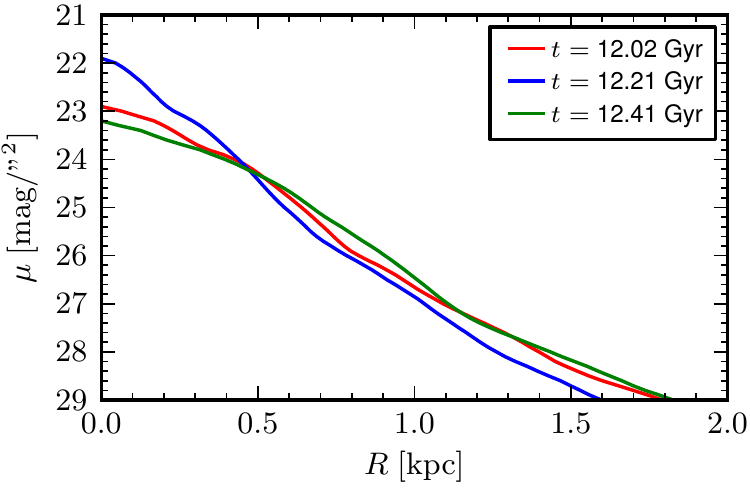}
\caption{Isophotal surface brightness profile in the B-band before
  (light gray, red in the online colour version), during (black, blue in the online colour version) and after (dark gray, green in the online colour version) the starburst in simulation B1.}
\label{fig:SB_2221}
\end{figure}

\begin{figure}
\includegraphics[width=0.47\textwidth]{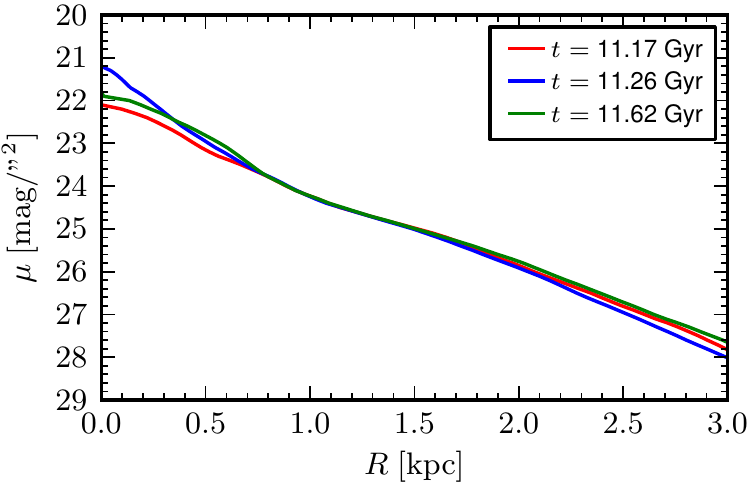}
\caption{Same as Figure \ref{fig:SB_2221} but for simulation B6.}
\label{fig:SB_3936}
\end{figure}

As shown in \S \ref{sec:density_profiles}, the stellar body of the
galaxy becomes more compact during a starburst, in agreement with
e.g. \cite{papaderos96a, micheva13a,micheva13b}. They draw their
conclusions by investigating the surface brightness profiles, usually
obtained using \emph{isophotal integration} due to the possible
irregular shapes of BCDs. This method was already used to produce the
colour profiles in \S \ref{sec:color}, and now we obtain surface
brightness profiles in a similar way. Figures \ref{fig:SB_2221} and
\ref{fig:SB_3936} show the B-band surface brightness (SB) for model B1
and B6 respectively, before (light gray, red in the online colour version), during (black, blue in the online colour version) and after (dark gray, green in the online colour version) the starburst

For the strong burst in B1 at $t\approx 12.2$ Gyr, we find an increase
in SB in the centre, by about 1~mag/arcsec$^2$, due to the bright,
newly formed stellar populations. Moreover, as shown in \S
\ref{sec:density_profiles}, the outer stars are moving inwards causing
the SB outside the inner $\sim 0.5$~kpc to decrease. This results in a
overall steeper surface brightness profile. For the more massive model
B6, we also find an increased SB in the centre and a decreased SB at
large radii, albeit less pronounced than in the low mass model
B1. This is because the burst is weaker relative to the average SFR of
the host galaxy (see Table \ref{tab:mergers}).

\begin{figure*}
\begin{minipage}{\textwidth}
\centering

\includegraphics[width=0.97\textwidth]{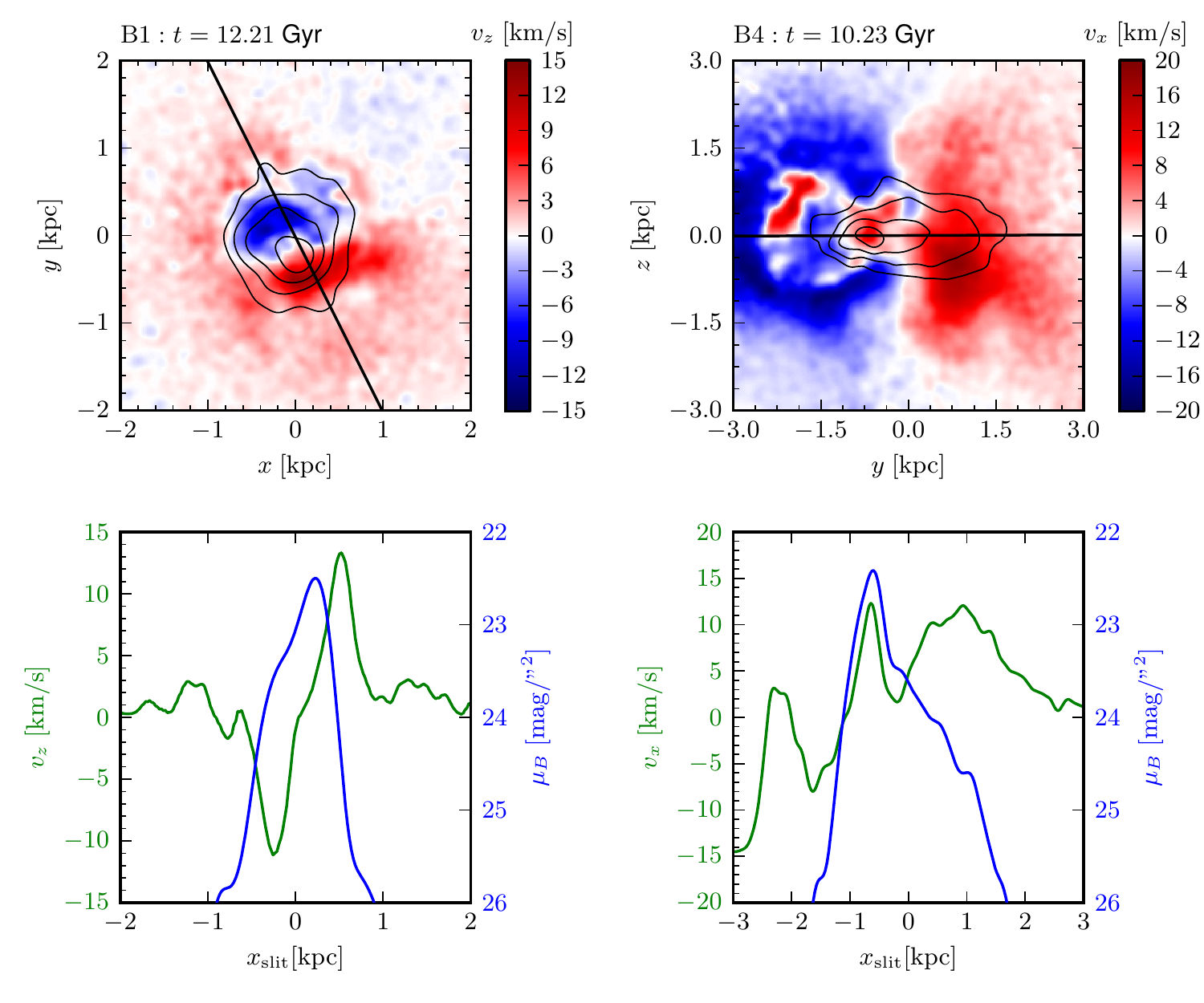}
\caption{Top: velocity field of the gas overlaid with the surface brightness contours in the B-band (top). Bottom: the line-of-sight velocity (gray, green in the online colour version) and the surface brightness (black, blue in the online colour version) along a slit put on the major axis for simulations B1 (left) and B4 (right).}
\label{fig:velocity_field}
\end{minipage}
\end{figure*}

\subsection{Subtypes} \label{sec:subtypes}
Based on the isophotal maps discussed in \S \ref{sec:isophotes}, we
can classify our models according to the classification scheme of
\cite*{loose86}. Since our bursting models clearly have an old stellar
population and we do not let two dwarf galaxies merge, we do not
expect i0 or iI,M BCDs to be found in our simulations.  

B1 and B6 have regular isophotes and a nuclear, albeit slightly
off-centre, starburst and could thus be classified as nE BCDs.  B5 has
regular isophotes as well but the starburst is very off-centre and/or
irregular, so we classify this model as a iE BCD. When viewing B3 and
B4 face-on, we see an irregular, off-centre starburst in a highly
irregular host. One would classify these as iI BCDs. However, when B3
and B4 are viewed edge-on, the starburst region can be significantly
off-centre, giving these models a flattened, cometary look. For this
viewing geometry, B3 and B4 would be classified as a iI,C. The same
classification applies to B2.

Based on these findings and the qualitative differences between the
models, we can propose an explanation and make predictions for the
different subtypes and their properties:
\begin{itemize}[leftmargin=0.5cm]
\item iE BCDs can be produced when the infalling gas cloud immediately
  triggers a burst in a not, or only slowly, rotating dwarf, as the
  dense gas will most likely not be centralized and have irregular
  shapes and the outer isophotes will have regular shapes (e.g. models
  B2, B5). Of course, if the star-forming region is significantly
  off-centre, the galaxy might be labelled as a cometary BCD.
\item nE BCDs on the other hand will more likely be found in secondary
  bursts in the same host galaxy (e.g. models B1, B6) as gas has
  re-accumulated at the galaxy centre and ignited in a new starburst.
  However, the supernova feedback from this central starburst can lead
  to star formation in dense clouds on the edges of supernova blown
  bubbles, resulting in more irregular star forming regions. Thus, a
  nE BCD may evolve into an iE BCD. 
\item iI BCDs are found in rotating host galaxies (e.g. models B3,
  B4). \cite{schroyen11} found that in simulated dwarf galaxies with a
  significant amount of rotation, star formation occurs continuously
  over the galactic body, in contrast to a more bursty, centralized
  star formation in dwarfs with slow (or no) rotation. This more
  dispersed star formation results in an irregular stellar body. So
  when the galaxy enters the starburst phase, the star forming region
  will most likely be off-centre and have an irregular shape. When
  looking at an iI BCD with off-centre star formation, it could be
  classified as an iI,C BCD, when viewed edge-on.
\end{itemize}
Our other simulations which produced a burst but which were not
explicitly discussed in this paper, all agree with this explanation
for the classification of the different subtypes as well.

There are indeed a number of observations that support this
picture. iI BCDs were found to be in general bluer \citep{gildepaz03},
more gas-rich, and more metal-poor \citep{zhao13} than nE/iE BCDs and
their {\sc HI} distribution is more clumpy rather than centralized
\citep{taylor94, vanzee98,thuan04b}. All of these findings can be
explained by the degree of rotation they have \citep{schroyen11}.
\cite{thuan04b} found that the nE and iE BCDs in their sample show no
regular rotation, while the two cometary BCDs in their sample are
rotationally dominated. At least 5 of the 7 iI,C BCDs investigated in \cite{sanchezalmeida13} are rotating. Observations have shown that some iE BCDs have their star forming regions organized in a ring \citep{gildepaz03},
supporting the idea of propagating star formaton and an evolution from
nE to iE BCD.

\cite{noeske00} found that the host galaxies of iI,C BCDs generally
have structural parameters that lie between those of BCDs and other
dwarfs. If our models are indicative for real life BCDs, the fact that
these are less compact than nE/iE BCDs can be explained in two
ways. Firstly, if star formation occurs very off-centre, there will
not be such a large increase in the central mass and thus the
gravitational potential will deepen less. This results in a less
compacthost galaxy. Secondly, iI,C BCDs are likely to be found in
rotating host galaxies, which are generally already more diffuse. On
the other hand, when looking at a flattened galaxy edge-on, the
surface brightness will drop more rapidly, resulting in a seemingly
more compact galaxy, as opposed to when we would view it face-on. We
thus propose that iI BCDs should be even more diffuse than iI,C BCDs,
and their hosts will have scale lengths and central surface
brightnesses in the parameter space occupied by dE and dIs. To the
best of our knowledge, no large sample study has been performed of the
structural parameters of the host of iI BCDs. The sample of
\cite{micheva13a,micheva13b} contains some iI BCDs, who do indeed tend
to lie in the parameter region of the dEs and dIs.

\subsection{Gas kinematics}

As the gas and stellar component of the galaxy becomes more
centralized, we expect that their rotational velocity will increase
due to conservation of angular momentum. This would explain the
observations by \cite{vanzee01}, who found that the {\sc HI} of BCDs
have steeply rising rotation curves. Figure \ref{fig:velocity_field}
shows the 2-dimensional gaseous line-of-sight velocity fields of B1
and B4 (edge-on view), left and right respectively, during the burst,
overlaid with the B-luminosity contours. We show these two quantities
along a slit placed along the major axis to mimic how this would be
detected observationally. For B1, we indeed find a steeply rising
rotation curve.

For B4, we see that the galaxy overall has solid body rotation around
the minor axis (the $z$-axis), but the kinematics of the starburst
region are very different \citep[as found by e.g.][]{ostlin04,
  koleva14}. In this model, a CGC has fallen in, coming from the right
side in Figure \ref{fig:velocity_field}. This impact will cause a
significant amount of gas in the white (red in the online colour version) region (with positive $v_x$) to
move to the black (blue in the online colour version) region (with negative $v_x$), causing a disturbance
in the solid body rotation pattern of the galaxy. Looking along the
$y$-axis (along which the gas cloud fell in), we will see a similar
disturbance while in the $z$-axis, no significant difference will be
found.

We can conclude that if the burst happens right after the gas cloud
fell in and depending on our viewing angle at the galaxy, the gas can
have very disturbed kinematics. In secondary bursts on the other hand,
like in B1, the gas will most likely show no disturbance.

\begin{figure}
\includegraphics[width=0.47\textwidth]{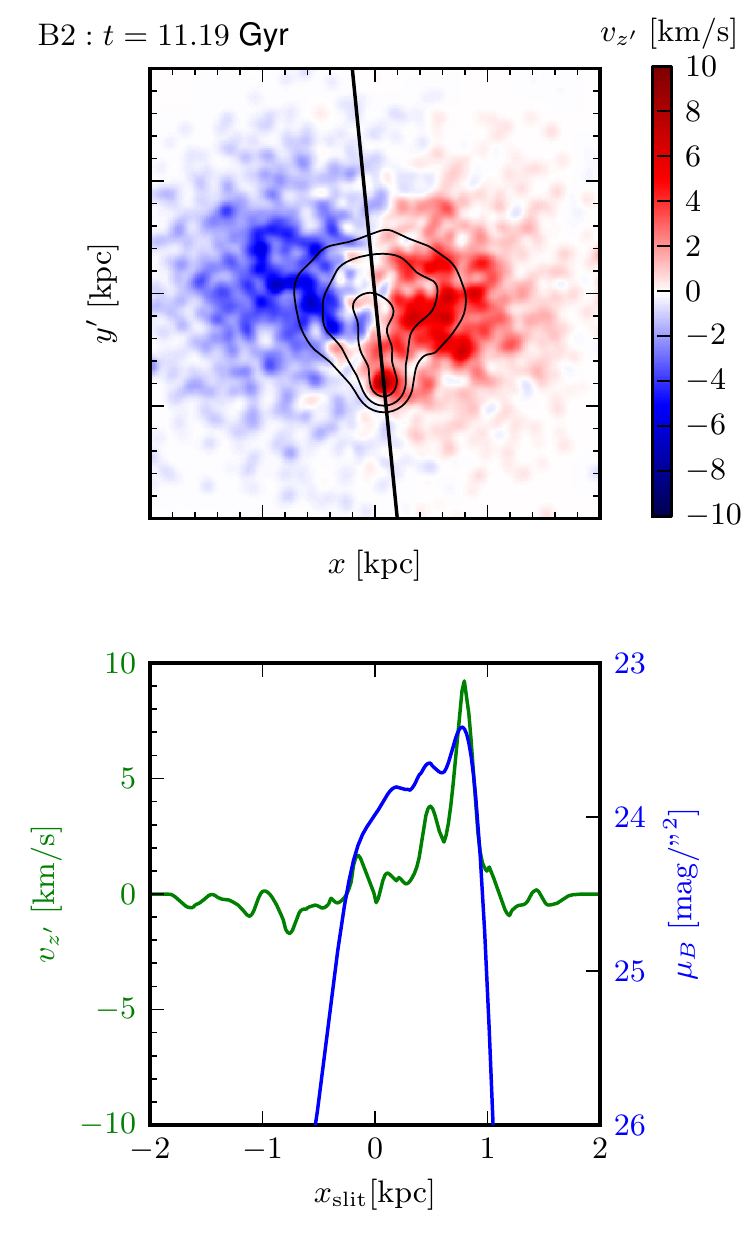}
\caption{Same as in Figure \ref{fig:velocity_field} but for the stars of simulation B2. Note that the simulation was rotated $\sim 50^\circ$ around the $x$-axis to show the effects more clearly.}
\label{fig:velocity_field_stars}
\end{figure}

\subsection{Stellar kinematics}

\cite{koleva14} found that the star forming regions of BCDs can have
kinematics that are very different from those of their host
galaxies. The top panel of Figure \ref{fig:velocity_field_stars} shows
the velocity fields of the stars of simulation B2. The radial velocity
and surface brightness along a slit along the major axis is shown in
the bottom panel. As we can see, the velocity of the gas cloud that
fell in results in a star forming region with different kinematical
properties than that of the host galaxy. Thus, our simulations are
able to reproduce the findings of \cite{koleva14}, at least for
simulations with a compact infalling gas cloud, as in B2.

\subsection{Evolutionary path}
\label{sec:evolution}
It has been long discussed what BCDs will look like when their
starburst has faded. Among others, \cite{papaderos96a} suggested that
postburst dwarf galaxies can not reside in the general dE or dI
population without a drastic change in their structural
parameters. \cite{vanzee01} state that a passive evolution to dE is
unlikely as it would require a significant loss of angular
momentum. They suggest that so-called compact dI
\citep[e.g.][]{vanzee00, patterson96} are candidate faded BCDs, since
they have similar scale lengths and have significant colour gradients.

\begin{figure}
\includegraphics[width=0.47\textwidth]{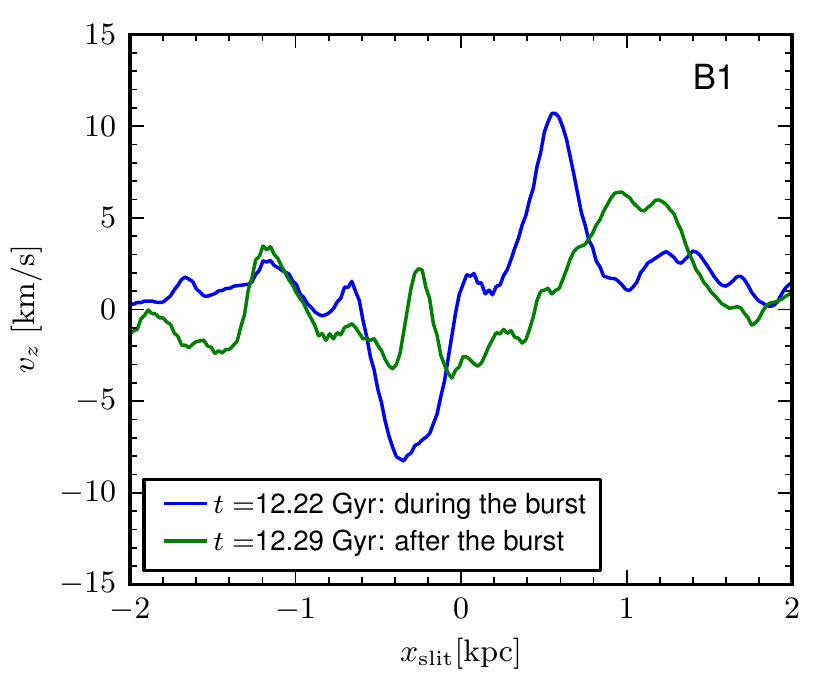}
\caption{Line of sight velocity of the gas along the same slit as in Figure \ref{fig:velocity_field} for simulation B1 during (black line, or blue in the online colour version) and after (gray line, or green in the online colour version) the starburst. }
\label{fig:velocity_field_after}
\end{figure}

Looking at our models, we find that generally postburst galaxies tend
to be more diffuse after the burst than during the burst, when the
galaxy would be classified as a BCD. Often, the galaxy is even more
diffuse than before the burst(s), placing it within the structural
parameter space occupied by dEs and dIs. Moreover, as the galaxy
re-expands after the burst, the steep rise of its rotation curve
disappears (as can be seen in Figure \ref{fig:velocity_field_after}), removing also this obstacle to a possible BCD-dI
transition. On top of that, the time-scale of a burst is much smaller
than the gas exhaustion time and the feedback following the starburst
is not strong enough to completely remove all the gas of the
galaxy. Thus BCDs that were dIs before the burst will be dIs again
after the burst.

\subsection{The role of feedback}

\begin{figure}
\includegraphics[width=0.47\textwidth]{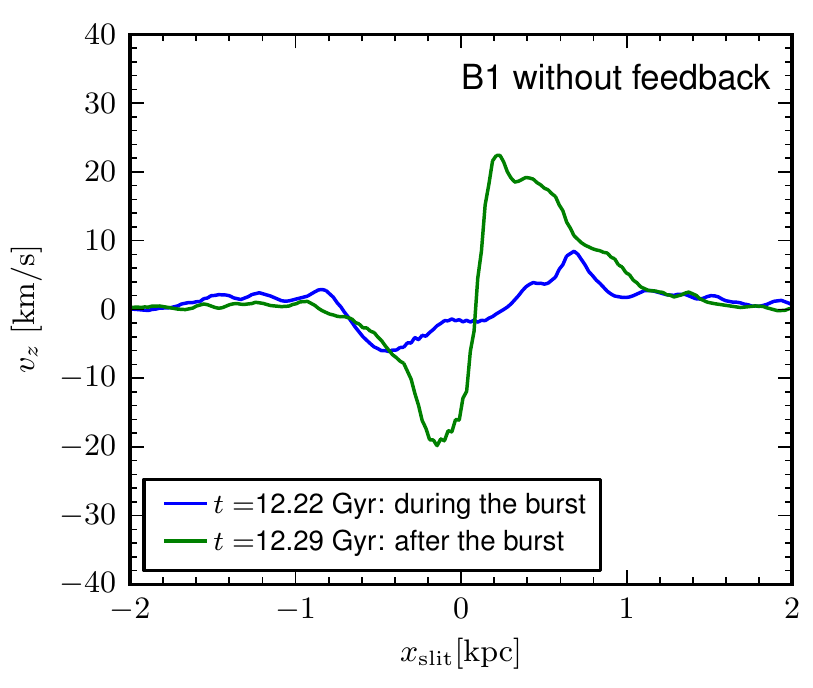}
\caption{ Same as in Figure \ref{fig:velocity_field_after}, but for a model with both star formation and feedback turned off after the burst. Note that the vertical scale here is different from that of Figure \ref{fig:velocity_field_after}.}
\label{fig:velocity_field_No_SN}
\end{figure}

Stellar feedback through supernovae and stellar winds is assumed to play a vital role in quenching the starburst. Furthermore, it seems that the strong feedback effects result in a diffuse post-burst galaxy, with shallow rotation curves and low stellar velocity dispersion, as discussed in paragraphs \ref{sec:density_profiles}, \ref{sec:evolution} and \ref{sec:vel_disp}. To confirm that this evolution is primarily due to feedback, we have rerun simulation B1, but with feedback turned off after $t=12.22$ Gyr (the time of the burst). Since there would be no way of quenching the starburst, we also turn off star formation.

There is now nothing to prevent the gas from collapsing further, resulting in a very deep gravitational potential, a very compact galaxy (in both gas, stars and dark matter), high velocity dispersion and a very steep rotation curve. The latter is shown in Figure \ref{fig:velocity_field_No_SN}.

\section{Summary}

We have run a large suite of numerical simulations of gas clouds
falling in on dwarf galaxies, using different galaxy and cloud masses
and sizes, and different orbital parameters. Our goal was to test
whether starbursts could be ignited this way and whether the resulting
galaxies could explain some of the properties of observed Blue Compact
Dwarfs.

Based on our simulations, we conclude that infalling gas can indeed
trigger a starburst in dwarf galaxies by significantly disturbing the
gas, especially in dwarf galaxies with $M_B \gtrsim -14$ mag. We
  found that gas cloud masses significantly below $\sim 10^7$
  M$_\odot$ do not trigger starbursts in the mass regime of the dwarfs
  simulated by us. If such clouds are rare, then this tells us that,
  although this is a viable way of triggering starbursts in dwarfs,
  this is not the triggering mechanism in the majority of the BCDs.
  Of course, less massive clouds could still trigger starbursts in
  dwarf galaxies with lower masses than the ones we simulated. The
starburst phase seems to be episodic, alternated with periods of
little star formation. Following the infall of metal-poor gas, the
overall gas metallicity of the galaxy drops. However, the stellar
metallicity can either increase or decrease, depending on the source
of the gas from which the stars are formed:~the pre-enriched galaxy or
the metal-poor infalling gas cloud. In most of our models, the stars
were formed from the old gas, resulting in metal-rich stars. Only in
models with compact infalling gas clouds can stars form from the newly
arrived gas, leading to more metal-poor newborn stars. Following the
impact of the gas cloud, the gas kinematics can become very
disturbed. Stars that form from kinematically distinct gas clumps tend
to be also kinematically peculiar when compared with the stellar body
of the galaxy.

As gas flows towards the galaxy centre to fuel the central starburst,
the gravitational potential deepens. In response, the whole body of
the galaxy becomes more compact and the rotation curve steepens. Once
gas is rapidly evacuated from the galaxy centre by supernova feedback,
the galaxy re-expands and becomes more diffuse again, flattening the
rotation curve. This removes some of the arguments against
evolutionary links between BCDs and dIs. 

As discussed in paragraph \ref{sec:subtypes}, we can attempt an
explanation for the physical differences between the subtypes of BCDs
in the classification scheme of \cite{loose86} within the scope of our
simulations:
\begin{itemize}[leftmargin=0.5cm]
\item nE and iE BCDs have host galaxies that are slowly (or not)
  rotating. nE BCDs can evolve into iE types as star formation
  propagates through the host galaxy.
\item iI BCDs have flattened host galaxies with strong rotation, viewed
  approximately face-on.
\item iI,C BCDs have flattened, strongly rotating host galaxy, viewed
  approximately edge-on.
\end{itemize}

\section*{Acknowledgements}

 We thank the anonymous referee for his/her prompt and constructive remarks
  that improved the content and presentation of the paper.  We thank
Volker Springel for making publicly available the {\sc Gadget-2}
simulation code. This research has been funded by the Interuniversity
Attraction Poles Programme initiated by the Belgian Science Policy
Office (IAP P7/08 CHARM). Annelies Cloet-Osselaer, Sven De Rijcke and
Bert Vandenbroucke thank the Ghent University Special Research Fund
(BOF) for financial support. Mina Koleva is a postdoctoral fellow of
the Fund for Scientific Research – Flanders, Belgium (FWO11/PDO/147).

\bibliographystyle{mn2e}
\bibliography{bibliography}

\label{lastpage}

\end{document}